\documentclass[prd,aps,preprintnumbers]{revtex4}%[twocolumn]%{revtex4}
\usepackage{epsfig,subfigure}
\usepackage{amsmath,amsfonts}

\usepackage{epstopdf}

\def \nn {\nonumber}
\def \be  {\begin{equation}}
\def \ee  {\end{equation}}
\def \bea  {\begin{eqnarray}}
\def \eea  {\end{eqnarray}}
\def \ba {\begin{eqnarray*}}
\def \ea {\end{eqnarray*}}
\def \bb  {\begin{thebibliography}}
\def \eb  {\end{thebibliography}}

\newcommand \bi [1] {\bibitem{#1}}
\def \lab #1 {\label{#1}}

\def \fracs #1#2 {\mbox{\small $\frac{#1}{#2}$}}

\def \bin #1#2 {{\left({#1}\atop{#2}\right)}}
\def \as {\relax\ifmmode\alpha_s\else{$\alpha_s${ }}\fi}

\def \al #1 {\frac {\as({#1})}{\pi} }
\def \ds #1 {\ooalign{$\hfil/\hfil$\crcr$#1$}}

\def \bi {\bibitem}

\def \O {\Omega}

\def\figscale#1#2{\pdfximage width#2 {#1.pdf}\pdfrefximage\pdflastximage}

\def\hepph  #1 {{\tt hep-ph/#1}}

\def\as{\ensuremath{\alpha_{s}}}
\def\a0{\alpha_0}

\def\vep{\varepsilon}

\def \ep{\epsilon}

\def \nn {\nonumber}

\def\bea {\begin{eqnarray}}
\def\eea {\end{eqnarray}}

\def\be {\begin{equation}}
\def\ee {\end{equation}}
\def\bi {\begin{itemize}}
\def\ei {\end{itemize}}

\begin{document}

%\preprint{YITP-SB-23-35}

\renewcommand{\thefigure}{\arabic{figure}}

\title{Soft photon theorem in  QCD with massless quarks}

\author{Yao Ma}
\email{yaomay@phys.ethz.ch}
\author{George Sterman}
\email{george.sterman@stonybrook.edu}
\author{Aniruddha Venkata}
\email{aniruddha.venkata@stonybrook.edu}

\affiliation{${}^*$Insitute for Theoretical Physics, ETH, 8093  Zurich, Switzerland\\
${}^{\dagger,\ddagger}$C.N. Yang Institute for Theoretical Physics and Department of Physics and Astronomy,
\centerline{Stony Brook University, Stony Brook, New York, 11794--3840, USA}
}
\date{\today}

\begin{abstract}
Working to all orders in dimensionally-regularized QCD,  
we study the radiation of a photon whose energy is much lower than that of external partons,
but much larger than the masses of some quarks.   We argue that 
the conventional soft photon theorem receives corrections at leading power in the photon energy, associated with soft virtual
loops of massless fermions.   
These additive corrections give an overall factor times the non-radiative amplitude that is infrared finite and real to all orders in $\alpha_s$.
Based on recent calculations of the three-loop soft gluon current, we identify the lowest-order three-loop correction.
\end{abstract}

\maketitle

%\section{Introduction}

Electromagnetic radiation in the scattering of charged particles through the strong interactions
 is described at low energy, $\omega(k)$, by the celebrated soft photon theorem,
which determines the first two powers of  $\omega(k)$ in terms of the non-radiative amplitude \cite{Low:1958sn,Burnett:1967km,Weinberg:1965nx,Gribov:1966hs,DelDuca:1990gz}.
 For all charged particles massive,
the amplitude to emit a photon with sufficiently small energy $\omega(k)$ and polarization $\epsilon(k)$ is given schematically by
\bea
M_3 (\{p_i\},k,\epsilon(k))\ =\  \sum_i\, \delta_i\, e_i\,  \frac{ p_i^\mu}{p_i\cdot k}\; \, 
\big [  \ep_\mu(k)\ -\ \left  (   k_\mu\epsilon^\nu(k) -\, \epsilon_\mu(k) k^\nu \right )\, O_\nu \left (p_i,k \right )  \big ]\, M_2(\{p_i\})\, ,
\label{eq:low-lo}
\eea
where $ M_2$ is the non-radiative amplitude, and where $\delta_i=+1$ ($-1$)
for particle $i$ with charge $e_i$ outgoing (incoming).  In the  correction term, $M_2$ is acted on by a momentum-dependent operation $O_\nu(p_i,k)$  \cite{DelDuca:1990gz,Sahoo:2018lxl,Krishna:2023fxg} that
does not change the power of $k$. 
Working in QCD, we will study soft photon production in pair creation processes: ${\rm singlet} \rightarrow f(p_1)+\bar f(p_2)$, for particles with electromagnetic charges $\pm e_f$. 
 We will be concerned
primarily with the leading power, but extensions of the reasoning below can be applied to the first power correction as well. 
We will always assume a fixed angle of order unity between the photon and either of the charged particles.

As extended by Gribov \cite{Gribov:1966hs} and Del Duca, \cite{DelDuca:1990gz}, Eq.\ (\ref{eq:low-lo}) holds for fixed-angle radiation so long as $\omega(k)\ll m$,  the
mass of any charged particle.   With this assumption, Eq.\ (\ref{eq:low-lo}) applies 
quite generally, and specifically for dimensionally-regularized perturbative amplitudes for QCD with massive quarks.  
 In this short paper,
we show that for pair creation or annihilation
in {\it massless} QCD, or when the photon energy exceeds the masses of some quarks, the relation Eq.\ (\ref{eq:low-lo}) 
receives corrections that  are of leading power, $1/\omega(k)$.   Specifically, we will show that for the pair production 
amplitude of  flavor $f$, keeping leading power only,
\bea
M_3^{(f)}(\{p_i\},k,\epsilon(k))\ =\  \left[ e_f\; +\; \Gamma_{\rm  EM}^{(f)}(\as) \left( \sum_{n=1}^{n_0} e_n \right )\right ]  \left[ \frac{p_1\cdot \epsilon(k)}{p_1\cdot k}\ -\  \frac{p_2\cdot \epsilon(k)}{p_2\cdot k} \right ] \, 
M_2^{(f)}(\{p_i\})\, .
\label{eq:low-ho}
\eea
Here, $\Gamma^{(f)}_{\rm EM}(\as)$, which we will refer to as the soft photon factorization constant, is an infrared finite expansion in the strong coupling.
It comes from loops of massless quarks (charges $e_n$), and begins at order $\as^3$.   The corresponding contribution when $\omega(k)\ll m$ appears in the subleading term in 
Eq.\ (\ref{eq:low-lo}).
This additive correction in Eq.\ (\ref{eq:low-ho}) is independent of the electromagnetic charges of the external particles, but is sensitive
to their color representations.  As a result of gauge invariance,  the
tensor structure of the correction matches that of the familiar form Eq.\ (\ref{eq:low-lo}).  In principle, $\Gamma_{\rm EM}^{(f)}$ may also be a function of the external momenta, but as we shall see, it is independent of momenta at lowest order in four dimensions.
We will extract the explicit value of $\Gamma_{\rm EM}^{(f)}$ at lowest nonvanishing order from the explicit calculation of the soft gluon current
in Refs.\ \cite{Chen:2023hmk} and \cite{Herzog:2023sgb} using a method we describe below.
A corresponding result
holds as well for massless quantum electrodynamics,  
 with an analogous $\Gamma_{\rm EM}$ beginning at $\alpha^3$.  
This gauge invariant contribution is additional to the classic leading terms in Eq.\ (\ref{eq:low-lo}), which have been derived in recent years for massless QED
from asymptotic symmetry considerations
\cite{Strominger:2017zoo}.

%\section{Soft photons and leading regions}

Our discussion starts with the identification of regions of loop momentum space in diagrams that contribute to 
the amplitude $M_3^{(f)}$ and that are leading in both the large scale of the process and the smaller scale of the photon energy.   
Infrared, mass-sensitive behavior arises only where loop momenta are pinched by propagator poles, called pinch surfaces in Ref.\ \cite{Sterman:1995fz}.

For the exclusive amplitude, $M_2^{(f)}(p_1,p_2)$, the pinch surfaces are uniquely identified by subsets of lines that either are on shell and collinear to 
external on-shell particles (jet subdiagrams) or have vanishing momenta (soft subdiagram), with  the remaining lines
off shell (hard subdiagram) \cite{Mueller:1979ih,Collins:1980ih,Sen:1981sd}. 
Straightforward power counting shows that singularities of the exclusive amplitude are logarithmic \cite{Sterman:1995fz}.  
Factorized forms of the exclusive amplitudes in terms of jet, hard, and soft functions, $M_2= H\, J_1\, J_2\, S$, are well-established in direct QCD
\cite{Collins:1980ih,Sen:1981sd,Korchemsky:1988hd,Dixon:2008gr,Feige:2014wja,Erdogan:2014gha,Ma:2019hjq}
 and soft-collinear effective theory \cite{Bauer:2001yt,Becher:2014oda}.  We
 will follow the matrix element conventions for
 the jet functions $J_i$ in Ref.\ \cite{Dixon:2008gr},  in which soft singularities are subtracted from the jet functions to avoid double counting with the soft function.

\begin{figure}[h]
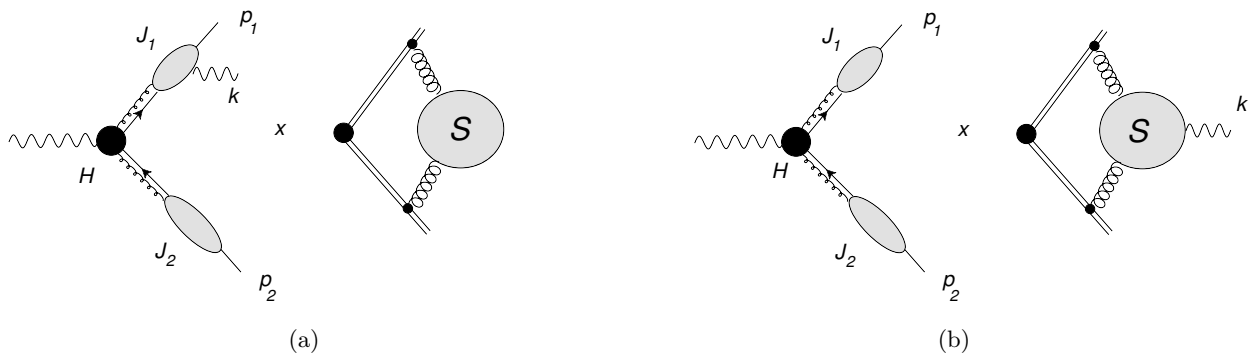

\centerline{\figscale{factored-full-jet}{6.6cm}\ \hspace{2.3cm} \figscale{factored-full-soft}{7.4cm}}

\medskip

\centerline{ (a) \hspace{8cm} (b) }
\caption{Leading regions for the soft gluon emission amplitude $M_3^{(f)}$, after factorization of soft lines from jets: (a) radiation from jets ($J_i$) associated with external lines, (b) radiation from fermion loops in
soft subdiagram, $S$.   $H$ represents the hard subdiagram.   The gluon lines in the figure represent arbitrary numbers of propagators connected
to the Wilson lines or hard subdiagram. The leading regions for the elastic process $M_2^{(f)}$ lack the photon, but are otherwise identical. \label{fig:leading-regions}}
\end{figure}

Turning to the radiative amplitude, $M_3^{(f)}$, because the photon energy is small compared to the momenta of the external charged particles, the leading,
${\cal O}(1/\omega(k))$ behavior can only arise in loop momentum configurations where the soft photon emerges either from (anti)quarks in the jet subdiagrams,
collinear to the external particles, or from soft
fermion loops in the soft subdiagram.   These contributions, which are additive, are illustrated in 
Figs.\ \ref{fig:leading-regions}a and \ref{fig:leading-regions}b, respectively.   In both cases, soft quanta, including those that carry
the momentum of the photon, factor from jet subdiagrams at leading power in $\sqrt{s}$.  Soft-jet-hard factorization,
which occurs in every leading region, applies to the radiative amplitude, $M^{(f)}_3$, just as for the exclusive amplitude, $M_2^{(f)}$.
In the factorized forms of both the exclusive and radiative amplitudes,   the soft function is given by the products of Wilson
lines in the directions $\beta_i \propto p_i$,  $P\, \exp [ \int_0^\infty  d\lambda (-ig_s)\beta_i\cdot A^{(R_i)}(\beta_i\lambda)]$, in the color representations $R_i$ of the external particles
\cite{Collins:1980ih,Sen:1981sd,Korchemsky:1988hd,Dixon:2008gr,Feige:2014wja,Erdogan:2014gha,Ma:2019hjq,Bauer:2001yt,Becher:2014oda}.    We note that the radiative soft function $S$ in Fig.\ 1b has additional pinch surfaces due to the
presence of the external photon, as we discuss below.

\begin{figure}[h]
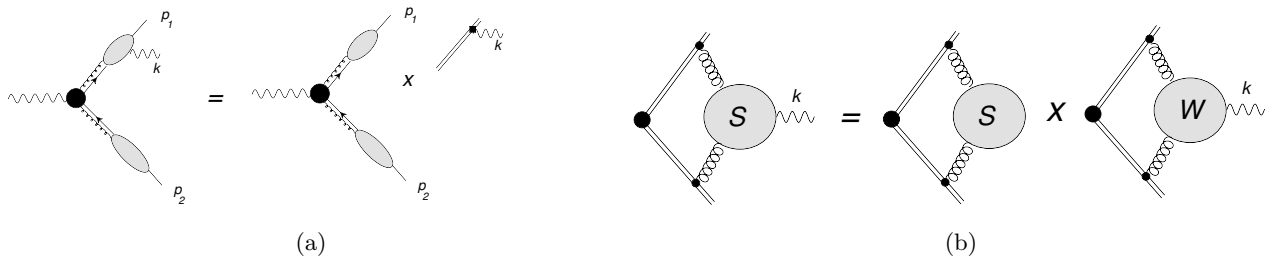

\centerline{\figscale{factored-jet}{6.6cm}\ \hspace{1.5cm} \figscale{factored-eikonal}{8.4cm}}

\medskip
\centerline{ (a) \hspace{8cm} (b) }
\caption{Factorization of the soft photon: (a) as eikonal factor from jet(s), (b) as part of a single radiative web, $W$ from the soft part, Eq.\ (\ref{eq:S-W-fact}).
The gluon lines  represent arbitrary numbers of soft or collinear connections in the general case.
\label{fig:soft-factor}}
\end{figure}

Based on the structure of both jet and soft loop emissions, the external soft photon enjoys further factorization properties. If it attaches to a jet, the factorization is illustrated by Fig.\ \ref{fig:soft-factor}a. The contributions of all quark loops within the jet cancel \cite{Gribov:1966hs,DelDuca:1990gz}, and the soft photon is sensitive only to the charge and velocity of the external particle. At leading power, this factorization gives the multiplicative eikonal factor represented by the double line and contributes to the $e_f$ term in Eq.\ (\ref{eq:low-ho}).

In contrast, if the external soft photon attaches to the soft subdiagram, the factorization is illustrated in Fig. 2b, where the soft photon contribution factors into the product of the soft function of the elastic amplitude $M_2^{(f)}$, times a diagram $W$ coupled by gluons to the product of Wilson lines. To avoid double counting, these Wilson lines do not couple to the photon field. As we will see below, this factorization comes about as a consequence of the exponentiation properties of products of Wilson lines (a single cusp for our case), resulting in the $\Gamma^{(f)}_{\rm EM}$ term in Eq.\ (\ref{eq:low-ho}).

%\section{Web analysis and soft photon factorization at leading power}

The exponentiation analysis of products of Wilson lines joined at a cusp \cite{Dotsenko:1979wb,Sterman:1981jc,Mukhi:1982bk,Gatheral:1983cz,Frenkel:1984pz,Laenen:2008gt}, see particularly Ref. \cite{,Mitov:2010rp},  
applies unchanged if we supplement the Lagrange density
of massless QCD by a vertex that couples quarks to a background electromagnetic field through the normal
quark electromagnetic currents.   If we take that field to be a plane wave of momentum $k$ with a corresponding
polarization vector, $\epsilon(k)$, the resulting soft function is an expansion of the general form, in $\varepsilon=2-D/2$ dimensions,
\bea
\bar S^{(f)}(\beta_1,\beta_2,k,\epsilon(k),\as,e,\varepsilon)
\ = \sum_{i,j} \left( \frac{\as}{\pi}\right)^i\, e^j\, \bar S_{(i,j)}^{(f)}(\beta_1,\beta_2,k,\epsilon(k),\varepsilon)\, ,
\label{eq:M-double-expand}
\eea
where $i$ and $j$ count the order in $\as$ and the unit electromagnetic charge, $e$ and therefore the loop order and the number of photon couplings, respectively. 
Our interest is entirely in the linear order in $e$, which by itself obeys the QED Ward identity, vanishing when $\epsilon^\mu(k) \propto k^\mu$.   
The complete sum, however, possesses  the same diagrammatic
exponentiation properties as QCD.   This is because the construction of the exponent
 depends only on analyzing attachments of the gluon field to the Wilson lines, and not at all on
the remainder of the diagrams to which
the gluons couple \cite{Dotsenko:1979wb,Sterman:1981jc,Mukhi:1982bk,Gatheral:1983cz,Frenkel:1984pz,Laenen:2008gt,Mitov:2010rp}.   

 The  result of this analysis applied to the singlet cusp is that the sum of all diagrams takes an
 exponential form in terms of so-called web diagrams, 
 which are constructed recursively as combinations of crossed and connected diagrams coupled to Wilson lines, with modified color factors.
In this case, the web functions themselves can be expanded
 in the electromagnetic charge as well as the strong coupling,
 \bea
 \bar S^{(f)}(\beta_1,\beta_2,k,\epsilon(k),\as,e,\varepsilon) \ =\ \exp \left [
 \int \frac{d^D l}{(2\pi)^D}\, \frac{\beta_1\cdot\beta_2}{l\cdot \beta_1\, l\cdot \beta_2}\ \frac{1}{l^2}\  
 \sum_{i,j} \, \left( \frac{\as(\mu,\vep)}{\pi}\right)^i\, e^j\, w^{(f)}_{(i,j)} \left( \{\beta_i\},l,\mu,k,\epsilon(k),\varepsilon \right) \right ]\, ,
\label{eq:web-expand}
\eea
where all loop integrals internal to the web diagrams $w^{(f)}_{(i,j)}$ have been carried out, leaving only the loop $l$ that joins the web with the cusp vertex.
Factoring an explicit $1/l^2$ as shown, the web functions are dimensionless.
For simplicity, and because we will limit ourselves to $j\le 1$,  we exhibit only a single photon momentum and polarization, representing  $j$ 
background photons.   Of course, the functions $w^{(f)}_{(i,0)}$ do not depend on $k$ and $\ep({k)}$.
 In the following, we refer to the function
$w^{(f)}_{(i,1)}$ in Eq.\ (\ref{eq:web-expand}) as the $i$th order radiative web function.

The web functions in Eq.\ (\ref{eq:web-expand}), which are integrated over their internal momenta at fixed $l$,
are renormalization-scale independent in the strong coupling \cite{Dotsenko:1979wb,Sterman:1981jc}, while the background photon field is unrenormalized.
Thus at fixed $j$, 
\bea
\mu{d\over d\mu}\;  \sum_{i} \, \left( \frac{\as(\mu,\vep)}{\pi}\right)^i\, e^j\, w^{(f)}_{(i,j)} \left( \{\beta_i\},l,\mu,k,\epsilon(k),\varepsilon \right)\ =\ 0\, .
\label{eq:mu-invar}
\eea
In addition, the web functions at every order are free of soft and collinear singularities
when subsets of lines have vanishing momenta or become collinear to the velocities $\beta_1$ or 
$\beta_2$.   This property \cite{Frenkel:1983di}, which follows from the web construction, as explained in Ref.\ \cite{Berger:2002sv}, is inherited by the focus of our interest, the linear term in the
electromagnetic charge, which is found simply by expanding the exponent of Eq.\ (\ref{eq:web-expand}),
\bea
\bar S_1^{(f)}(\beta_1,\beta_2,k,\epsilon(k),\as,e,\vep) &=&  e^{
 \int \frac{d^D l}{(2\pi)^D}\, \frac{\beta_1\cdot\beta_2}{l\cdot \beta_1\, l\cdot \beta_2}\, \frac{1}{l^2}
 \sum_{i=0}^\infty \, \left( \frac{\as(\mu,\vep)}{\pi}\right)^i\, w^{(f)}_{(i,0)} \left( \{\beta_i\},l,\mu,\varepsilon \right) }
\nn\\[2mm]
&\ & \hspace{5mm} \times\
 \ \int \frac{d^D l}{(2\pi)^D}\, \frac{\beta_1\cdot\beta_2}{l\cdot \beta_1\, l\cdot \beta_2}\ \frac{1}{l^2}\  
 \sum_{i=3}^\infty \as^i(\mu,\vep) \, e\, w^{(f)}_{(i,1)} \left( \{\beta_i\},l,\mu,k,\epsilon(k),\varepsilon \right)\, .
\label{eq:S-W-fact}
\eea
The first factor on the right is the unrenormalized soft function for the non-radiative QCD  process \cite{Dixon:2008gr},
 given by the exponential of the sum of pure QCD webs, $w^{(f)}_{(i,0)}$ in Eq.\ (\ref{eq:web-expand}).   The renormalization of integral $l$ for the $w^{(f)}_{(i,0)}$ webs in
 the exponent defines the soft function for $M_2^{(f)}$ \cite{Dixon:2008gr}.
The second factor in Eq.\ (\ref{eq:S-W-fact}) has all the dependence on $k$ and $\epsilon(k)$. Its integral over $l$ is ultraviolet finite.
 It satisfies the QED Ward identity independently, and hence must take the form,
\bea
 \int \frac{d^D l}{(2\pi)^D}\, \frac{\beta_1\cdot\beta_2}{l\cdot \beta_1\, l\cdot \beta_2}\ \frac{1}{l^2}\  
 \sum_{i=3}^\infty \as^i(\mu,\vep) \, e\, w^{(f)}_{(i,1)} \left( \{\beta_i\},l,\mu,k,\epsilon(k),\varepsilon \right)
 \ =\  
\Gamma_{\rm  EM}^{(f)}(\as) \left( \sum_{n=1}^{n_0} e_n \right ) \left[ \frac{\beta_1\cdot \epsilon(k)}{\beta_1\cdot k}\ -\  \frac{\beta_2\cdot \epsilon(k)}{\beta_2\cdot k} \right ]\, ,
\label{eq:gamma-web}
\eea
where in the prefactor, $\Gamma^{(f)}_{\rm EM}(\as)$, any residual dependence on momenta is suppressed.
We also drop dependence of 
$\Gamma^{(f)}_{\rm EM}(\as)$ on $\vep$ here, because, as we  argue below,  both the integrand and the integral on the left-hand side of this expression
are finite in four dimensions.

Using Eq.\ (\ref{eq:gamma-web}) in Eq.\ (\ref{eq:S-W-fact}), renormalizing the soft function and multiplying by the jet and hard functions, we derive the second term in the soft photon theorem for massless QCD,
Eq.\ (\ref{eq:low-ho}). 
 This is the primary result of our analysis.  We emphasize at the outset that the $1/\omega(k)$ behavior of the right-hand side is entirely due to loops of massless quarks.  For photon
 energies below the scales of nonzero quark masses,
the coupling of a photon to the corresponding quark loop is suppressed by the ratio $\omega(k)/m$, and appears
in the power correction term in Eq.\ (\ref{eq:low-lo}).   A study of the turnover between these two regimes will certainly be of interest.

%\section{The radiative web at lowest and higher orders}

\begin{figure}[h]
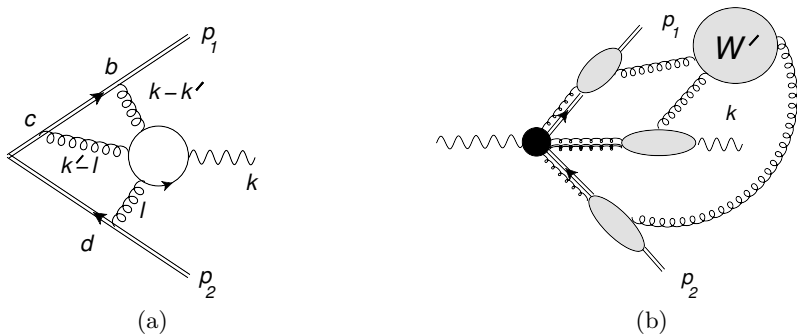

\centerline{\figscale{loop-3g-gamma}{3.4cm} \hspace{2.1cm} \figscale{general-pinch-order-k}{4.83cm}}
\centerline{(a) \hspace{60mm} (b)}
\caption{(a) Representative $\as^3$ quark-loop photon emission diagram for the soft function
of the radiative pair creation process, with assigned gluon momentum.  Loop momentum $l$ joins the cusp and the web subdiagram
as in Eq.\ (\ref{eq:S-W-fact}).   Color indices $b$, $c$ and $d$, which link the
fermion loop and Wilson lines, are discussed in the text.  
(b) Most general pinch surface for individual diagrams that contribute to the radiative web function, $w^{(f)}_{(i,1)}$.   Here, the double straight lines represent
the Wilson lines and the double gluon lines, gluons carrying physical polarization.  All other gluon lines represent arbitrary numbers of connections.
In the sum of web diagrams, all collinear singularities associated with jet subdiagrams on the $p_1$ and 
$p_2$ lines cancel.
\label{fig:lo-radiative}
}
\end{figure}

The  web diagrams in Eq.\ (\ref{eq:gamma-web}) are represented schematically
on the right of Fig\ \ref{fig:soft-factor}b.   Because of color and charge conjugation
invariance, the lowest nonvanishing order for the soft-photon web  is $\as^3$, illustrated in Fig.\ \ref{fig:lo-radiative}a.
The three-loop eikonal diagram shown in Fig.\ \ref{fig:lo-radiative}a and the general
radiative web functions on the right of Fig.\ \ref{fig:soft-factor}b
are proportional to $1/\omega(k)$ after the integral over $l$ in Eq.\ (\ref{eq:gamma-web}).  This behavior is characteristic of the ``soft" region where all components of momentum
in the three loops are of order $\omega(k)$. This can be verified readily by standard power counting for soft singularities, as described, for example, in
 Ref.\ \cite{Sterman:1995fz}.  

 To verify that the coefficient $\Gamma_{\rm EM}^{(f)}$ is
infrared finite, we must confirm that the complete radiative web has no infrared divergences in four dimensions,
even though its individual diagrams have many leading power pinch surfaces.
  In Fig.\ \ref{fig:lo-radiative}b,
we show a typical leading pinch surface for a radiative web.   Subsets of virtual lines form jet subdiagrams (shaded) collinear to
the $p_1$, $p_2$ or $k$ directions, with other lines, labelled $W'$, soft in all components relative to the momenta of the jet subdiagrams.
Note that the soft subdiagram $W'$ is connected to jets only through gluon lines.
Momentum configurations where a fermion loop appears in both jet and soft subdiagrams are always suppressed compared to leading power \cite{Sterman:1995fz}.
The fermion loop to which the photon connects must therefore be entirely in one of the $k$, $p$ or $p'$ jets, or in the soft function.
Individual graphical contributions to $\Gamma_{\rm EM}^{(f)}$ inherit poles up to $1/\ep^5$ for diagrams at ${\cal O}(\as^3)$
from the general configurations illustrated by Fig.\ \ref{fig:lo-radiative}b, and every higher order of $\alpha_s$ increases the power of $\vep$ by two.

We first show how, after combining diagrams, all leading-power singularities associated with nontrivial $p_1$ and $p_2$ jets cancel.
In diagrams like Fig.\ \ref{fig:lo-radiative}a, and indeed at any order, if the momentum of the quark loop reaches a scale much larger than $k$
and is collinear to $p_1$ or $p_2$, the diagrams found by connecting
photon $k$ to the loop in all ways cancel by the QED Ward identity \cite{Gribov:1966hs,DelDuca:1990gz}.
At high orders, leading pinch surfaces like those shown in Fig.\ \ref{fig:lo-radiative}b do occur, in which
the $p_1$ and $p_2$ Wilson lines are accompanied by collinear jet subdiagrams connected to the remainder of
the on-shell diagram by soft gluon lines.   As noted above, however,
such singularities cancel in the sum of web diagrams at any order.

In the remaining leading pinch surfaces in Fig.\ \ref{fig:lo-radiative}, the entire fermion loop
is part of a virtual jet of lines collinear to the photon.   This is an unusual jet, however, because although it originates at the Wilson line as a set of collinear gluons, its
entire momentum reaches the final state as a color-singlet photon \cite{Sterman:1977cj}.   Such a jet can be factored from the Wilson line, in
the same manner as the jets of the form factor.   At the same time, soft gluons also can be factored from this jet 
\cite{Collins:1980ih,Sen:1981sd,Korchemsky:1988hd,Dixon:2008gr,Feige:2014wja,Erdogan:2014gha,Ma:2019hjq,Bauer:2001yt,Becher:2014oda}, coupling only
to a Wilson line in adjoint representation and in the direction of momentum $k$.   These soft gluons would multiply a gauge invariant jet subdiagram that 
describes a gluon (with physical polarization) at short distances yet evolves without radiation into a single photon final state.   
Such a matrix element would vanish by color conservation.  This ensures the cancellation of infrared poles
associated with gluons much softer than $\omega(k)$ at such leading pinch surfaces, and consequently the infrared finiteness of $\Gamma_{\rm EM}^{(f)}(\as)$.

Given  the complexity of the diagrams involved, with poles up to $1/\ep^5$ in
individual diagrams, the lowest nonvanishing order is already a strong test of the reasoning above.
Remarkably, very recent calculations in massless QCD include precisely this
quantity, up to an overall shift in color factor.   References \cite{Chen:2023hmk} and \cite{Herzog:2023sgb} both
calculate the diagrams  of the type shown in Fig.\ \ref{fig:lo-radiative}a, but with an external gluon  rather than photon,
as part of the much larger calculation of the soft gluon current.  For an external
gluon, these diagrams have poles in dimensional regularization up to $1/\vep^6$ in
their dominant color structure.  At this order, however, a maximally symmetric color factor appears, equal to $(1/4!)\, {\rm Tr}[T^a T^b T^c T^d + {\rm permutations} ]$,
with generators $T^i$ in the fundamental representation.  (Comparing to Fig.\ \ref{fig:lo-radiative}a, the external gluon would have color index $a$.)
This color structure is present in every diagram like Fig.\ \ref{fig:lo-radiative}a, with three gluons connecting the
fermion loop to the Wilson lines, and only in those diagrams.
  
We now observe that the subset of gluon emission diagrams that provide the maximally symmetric color tensor are in one-to-one correspondence to diagrams
in the {\it full} set of diagrams for radiation of a photon.   
After using charge conjugation invariance, the surviving color factor from each quark loop diagram for photon emission
is proportional to the symmmetric tensor $(1/4)d^{bcd}=(1/2){\rm Tr}[T^bT^cT^d+T^dT^cT^b]$.  For both gluon and photon emission,
the result of the quark loop color trace is contracted with the same color generators, $T_{( R_f)}^{e}$, $e=b,c,d$, on the Wilson lines, whose relative order
depends on the diagram in question, but whose product is invariant under permutations because of the symmetry of the traces.
The sets of gluon- and photon-radiative diagrams are thus related by the ratio of their respective color factors and the substitution of an electromagnetic charge $e_n$
for one power of  $g_s$.   This procedure gives for the radiative coefficients with external quarks of flavor $f=F$,
\bea
\Gamma_{\rm EM}^{(F)}\ =\  -\, \left(\frac{\as}{\pi}\right)^3 \, C_F^{(3)}\;
\left ( -\frac{\zeta_2}{2} \, +\, \frac{\zeta_3}{6}\, +\, \frac{5\zeta_5}{6} \right )   \, ,
\label{eq:Gamma-result}
\eea
where $C_F^{(3)}$ is the cubic Casimir of the $SU(n)$ Lie algebra, $C_F^{(3)}=d^{bcd}T_F^bT_F^cT_F^d$ in the fundamental representation.
The corresponding coefficient vanishes for the gluon, $T_F\to T_A$.   Analogous reasoning shows that the same 
formula holds for massless QED, with the substitutions $C_F^{(3)} \to 4$ and $\as \to \alpha$.   Unlike the familiar soft photon
theorem terms in Eq.\ (\ref{eq:low-ho}), the $\Gamma_{\rm EM}^{(f)}$ contribution depends directly on the color representations of external particles $f,\bar f$, rather than on their electromagnetic charges.

In summary, for the $\as^3$ diagrams of Fig.\ \ref{fig:lo-radiative}a and at higher orders, the integrals that determine
the soft photon factorization constant are free of subdivergences, giving $1/\omega(k)$ times
an infrared finite number.   As a result, all logarithms in the expansion of $\Gamma_{\rm EM}^{(f)}$ can only have arguments depending on $\omega(k)/\mu$.  Then
the renormalization scale invariance of the web, Eq.\ (\ref{eq:mu-invar}), implies that
the scale of the running coupling in $\Gamma^{(f)}_{\rm EM}$  is naturally chosen as $\omega(k)$. 
 Indeed, in the order $\as^3$ results of Refs.\ \cite{Chen:2023hmk,Herzog:2023sgb}, at higher orders in the $\vep$ expansion, logarithms
appear with argument $(p_1\cdot k)(p_2\cdot k)/(p_1\cdot p_2\mu^2) \sim (\omega(k)/\mu)^2$.

Another  feature of the explicit form in Eq.\ (\ref{eq:Gamma-result}) is that $\Gamma^{(f)}_{\rm EM}(\as)$ is a real number in four
dimensions.  This may appear surprising, given that
intermediate states found by cutting diagrams like Fig.\ \ref{fig:lo-radiative}a can clearly go on-shell.   An example in the figure would be
a state consisting of two on-shell eikonal lines and gluon $k'-l$.   In the region that gives this cut, gluon $k-k'\equiv k_1$ must 
satisfy $p_1\cdot k_1 =p_1^+k_1^-=0$, where we take $p_1$ in the light-cone plus direction.   In this region of momentum
space, we can construct a path that passes through gluon $k_1$ and then through a sequence of vertices and lines, always
against the flow of positive minus momentum.  This sequence  of lines avoids the photon line, which carries plus momentum
out of the diagram.  It must eventually reach either the $p_1$ or $p_2$ eikonal lines, along which it can be completed back to the
$k_1$ line as a loop.   By construction, the plus momentum flowing around this loop has poles only in the upper half plane.
As a result, we may deform the plus momentum of this loop to any large scale, and all lines in the loop will be either
far off-shell or will be absorbed into the $p_1$ jet.    But we have seen that both at order $\as^3$ and beyond,
such momentum configurations do not contribute at order $1/\omega(k)$.  A construction of this kind can be carried out
for any pole on an eikonal line in any diagram contributing to $\Gamma^{(f)}_{\rm EM}$, and therefore for any on-shell intermediate state.  Our factorization constant is thus real to all orders.

%\section{Summmary, massive quarks and conclusion}

Equation (\ref{eq:low-ho}) is of interest in its own right as a result in massless QCD and related
gauge theories.   Since it is well-defined at zero quark mass, it should also hold approximately for masses
small compared to the photon energy.   This 
suggests that the coefficient of the $1/\omega(k)$ behavior of the amplitude for soft photon emission changes with
the photon energy.  For $\omega(k)$ small enough, it is given by the classic formula, Eq,\ (\ref{eq:low-lo}) in terms of the charges of external
particles, but as the energy increases past the masses of charged fermions, new $1/\omega(k)$ contributions appear.
Indeed, to capture the complete $1/\omega(k)$ asymptotic behavior at very high energies for the exclusive process, 
we can take the number of massless quark flavors $n_0$ to count all quark masses less than $\omega(k)$.
This quantitative but formal result may have indirect phenomenological consequences.  It may, for example, provide a link
between perturbative analysis of amplitudes and the nonperturbative model analyses \cite{Shuryak:1989vn,Botz:1994bg,Hatta:2010kt,Kharzeev:2013wra,Wong:2014ila}
that have been proposed to explain the apparent 
excess of relatively soft photon emission in a variety of experiments \cite{Chliapnikov:1984ed,SOPHIEWA83:1992czx,DELPHI:2005yew}.   Clearly, at the perturbative
level, it will be interesting to explore possible implications of this analysis for the scattering of bound states, both charged and neutral. 
Other  interesting generalizations of these results will be to  more external colored particles \cite{Almelid:2015jia,Ma:2019hjq,Agarwal:2021ais}, more external photons, and
the next power in photon momentum.  

%\acknowledgments
We thank Charalampos Anastasiou, Franz Herzog, Eric Laenen, Lorenzo Magnea and Bernhard Mistlberger for very helpful discussions. This work was
supported by the National Science Foundation, 
grants PHY-1915093 and PHY-2210533.

\end{document}